\documentclass[showpacs,preprintnumbers,amsmath,amssymb,10pt]{revtex4}
\usepackage{amssymb}

\usepackage{amsmath}
\usepackage{amsfonts}
\usepackage{bm}

 \numberwithin{equation}{section}

 \newcommand{\R}{\mathbb{R}}
  \newcommand{\C}{\mathbb{C}}
 \newcommand{\N}{\mathbb{N}}

 \newcommand{\h}{\mathbb{H}}
 
 \newcommand{\Z}{\mathbb{Z}}
 \newcommand{\four}{\mathcal{F}}

 \newcommand{\norm}[1]{\left\Vert#1\right\Vert}

\begin{document}

\title[LQG black hole entropy mathematics]{The entropy of large black holes in loop quantum gravity:
\\ A combinatorics/analysis approach}

\author{X. Cao}%
 \email{xiangyu.cao@polytechnique.edu}
\affiliation{%
Ecole Polytechnique\\
Route de Saclay, 91128 Palaiseau, France
}%

\author{A. Barrau}%
 \email{aurelien.barrau@cern.ch}
\affiliation{%
Laboratoire de Physique Subatomique et de Cosmologie, UJF, INPG, CNRS, IN2P3\\
53, avenue des Martyrs, 38026 Grenoble cedex, France
}%

\date{\today}
\begin{abstract}
The issue of a possible damping of the entropy periodicity for large black holes in Loop Quantum Gravity is highly debated. Using a combinatorics/analysis approach, we give strong arguments in favor of this damping, at least for prescriptions where the projection constraint is not fully implemented. This means that black holes in loop gravity exhibit an asymptotic Bekenstein-Hawking behavior, provided that a consistent choice of the Immirzi constant is made.
\end{abstract}

\maketitle

\section{Introduction}

Loop Quantum Gravity (LQG) provides a consistent framework (see \cite{lqg_review} for introductory reviews) to perform a non-perturbative and background independent quantization of General Relativity (GR). It has now been realized that many different approaches --covariant quantization, canonical quantization and quantum geometry-- converge to the very same LQG theory \cite{rovelli0}. As far as applications to physical systems are concerned, the main successes of the model are unquestionably cosmology and black holes. In the cosmological sector, LQG was shown to be very effective in regularizing the Big Bang singularity and in naturally setting the initial conditions for inflation to occur (see \cite{lqc_review} for introductory reviews). In the black holes sector, it has provided a detailed framework to precisely compute the entropy in a fully quantum setting. Since the pioneering works (\cite{rovelli1}, \cite{ash}), many studies were devoted to the critical issue of the entropy of black holes in LQG (see, {\it e.g.} \cite{articles_bh}).

Basically, the idea is to use an isolated horizon as an inner boundary of the considered manifold. For a given area $A$ of a Schwarzschild black hole horizon, the
physical states arise from a punctured sphere whose punctures carry quantum labels (see, {\it e.g.}, \cite{diaz1} for an up-to-date detailed review and \cite{diaz2} for a detailed analysis). For small black holes, the striking feature derived in this approach is that, in addition to its linear growth as a function of $A$, the entropy displays an effective ``staircase'' behavior due to a constant periodicity basically independent of the smearing $\delta A$. Many works were devoted to the study of this key feature (see, {\it e.g.}, \cite{articles_bh}). The main question, which has been intensely debated, is to understand whether this crucial behavior still holds for large black holes. Several non-conclusive arguments were given either in favor or against a persistence of the periodicity. In this article, we establish that it is damped, at least when the ``projection constraint'' is not applied.

Mathematically, this translates into the following combinatorics problem: let $$T(x) = \sum_{k=1}^{\infty} \delta(x - \sqrt{k(k+2)}),$$
and let $S = T + T\star T +T\star T\star T+ \dots$ ($\star$ denotes the convolution). What can be said about $S$? Numerical investigations for 
(\cite{diaz2}) of $S$ at small values of $x$ ($x\sim 10^2$) suggest that the exponential growth of $S$ is modulated by a given ``periodicity''. We will show,
in the framework of distribution theory (or the theory of generalized functions), 
that for any nonnegative test function with compact support $\zeta(x)\in \mathcal{D}(\R)$, $$\langle S,\zeta(x-a) \rangle
=k\exp(\gamma a)(1+o(1)), a\rightarrow +\infty$$ where $\gamma$ is a positive constant that will be 
defined explicitly(see (\ref{def_gamma})) and $k$ depends on $\zeta$. This means that the ``periodicity'' actually damps out for large $a$ values and that the LQG black hole entropy (provided the appropriate choice of the Immirzi constant) has the same macroscopic behavior than predicted by Hawking and Bekenstein.

\section{Definition of the problem}\label{definition}

Throughout all this discussion, let $j_k = \sqrt{k(k+2)}$ for $k = 0, 1, 2, \dots$. Let us define the 
\textit{generating distribution} 
\begin{equation}\label{generateur}
T = \sum_{k = 1}^ \infty \delta_{j_k},
\end{equation} 
where $\delta_{a}(x) = \delta(x - a)$.
We shall also consider its better-behaved approximations: for
$t\geq 0$,
\begin{equation}\label{generateurs}
T_{t} = \sum_{k = 1}^ \infty \delta_{j_k}e^{-t j_k} = e^{-tx}(\sum_{k = 1}^ \infty \delta_{j_k}), 
\end{equation} 
so that $T = T_0$. 

As $T_t$ is supported inside $(j_1 - \epsilon, +\infty)$
for any $\epsilon > 0$, one can define its convolution with itself. We call 
$T^l$ the convolution of $l$ T's ($l = 1, 2, 3, \dots$), and, similarly for $T_t$. It is straightforward to
see that $T_t^l$ is supported inside $(l j_1 - \epsilon, \infty)$. It should be noted that the lower limit tends to infinity when $l\rightarrow +\infty$.
As a consequence, The infinite sum 
\begin{equation}
S_t := \sum_{l=1}^\infty T_t^l, t\geq 0
\end{equation}
is well-defined as a distribution. We define $S = S_0$ which is the object of our study. We need to 
understand its behavior in the interval $[A - a, A]$, \textit{for large $A$'s and with $a\sim O(1)$}.
The physical motivation for this will be discussed later in Section \ref{discussion}.\\

\subsection*{Fourier-Laplace analysis-Basics}
The main analytic tool used in this study is the Fourier-Laplace transform (or holomorphic 
Fourier transform, also known as generating function method). It has already been used by several authors (\cite{articles_bh}). 
This section aims mainly at defining notations for the refined mathematical analysis of Section \ref{analysis}.

Let $f$ be the holomorphic Fourier transform of $T_0$, defined in the lower upper-plane:
\begin{equation}
f(z) = \langle T_0, e^{-izx} \rangle , z\in \h = \{p - it: p \in \R, t>0\}.
\end{equation}
From the definition of $T_0 = \sum_{k=1}^\infty \delta_{j_k}$, we have immediately
\begin{equation}
f(z) = \sum_{k=1}^\infty e^{-ij_k z}.
\end{equation} 
Note that for $t>0$, the function $p\mapsto f_n(p - it)$ is the Fourier transform (on the real line) of $T_t$. 
A first bound on $f$ can be given by  
\begin{equation}\label{easiest_est}
|f(p-it)|\leq \sum_{k=1}^\infty e^{- j_k t}.
\end{equation} 
Therefore, as $t\rightarrow+\infty$, $f(p-it)\rightarrow 0$ uniformly with respect to $p\in \R$. 
In particular, let $\gamma$ be the 
real number that satisfies 
\begin{equation}\label{def_gamma}
\sum_{k=1}^\infty e^{- j_k \gamma} = 1,
\end{equation}
then $|f(z)| < 1$ whenever $Im(z) < -\gamma$. Thus, in this region, the sum 
$ \sum_{l=1}^\infty f^l(z)$
converges to the holomorphic function 
\begin{equation}
F(z) = \frac{f(z)}{1-f(z)},
\end{equation}
while in the whole half plane $\h$, $F(z)$ defined by the above equation
is a meromorphic function (whose poles are roots of $1-f(z)=0$). 
This corresponds to the fact that $S=S_0$ is not a well-tempered distribution (actually, it grows exponentially). However, for any fixed $t>\gamma$, it is easy to see that 
this convergence can also be seen as a convergence in the space of tempered distributions (being uniform on $\R$). Applying the inverse Fourier transform (which is a continuous mapping of the space of well-tempered distributions to itself) on both sides yields:
\begin{equation}
\begin{split}
\four^{-1}(F(p-it))& =  \four^{-1}(\sum_{l=1}^\infty f(p-it)^l) \\
& =   \sum_{l=1}^\infty \four^{-1}(f(p-it)^l) \\
& =   \sum_{l=1}^\infty  T_t^l =  S_t.\\
\end{split}
\end{equation}
This establishes that for any $t>\gamma$, $S_t$ is well-tempered and its Fourier transform is $p\mapsto F(p-it)$.
In this sense $F(z)$ is the holomorphic Fourier transform (Fourier-Laplace transform) of $S$.

\section{Asymptotic Analysis}\label{analysis}
\textbf{Notations:} For $x\in\R$, $x \text{ mod }2\pi$ is the modulo 
map that takes value in $(-\pi,\pi]$. For $z\in\C$, let
$\norm{z}=\max(|Re(z)|, |Im(z)|)$, and $R(z,r)$ denote
the ``$r$-ball'' in this norm, that is, a rectangle
$2r\times 2r$ centered at $z$ (boundary excluded). Notations involving ``$\mathrm{dist}$'' and like $B(z,r)$ are to be considered in the ordinary Euclidean meaning.

\subsection{Isolation of the dominant contributions}

Let $a=p-it$ with $p,t\in\R, t\in(0,\gamma+1) $. We estimate the
quantity 
\begin{equation}\label{esti0}
\begin{split}
Re(1- f(a)) = & 1 - \sum_{k=1}^{\infty} e^{-j_k t}\cos(j_k p) 
            =  \sum_{k=1}^{\infty} \left( e^{-j_k\gamma} - e^{-j_k t}\cos(j_k p) \right) \\
            = & -\sum_{k=1}^{\infty} (e^{-j_k t }- e^{-j_k\gamma}) + \sum_{k=1}^{\infty} e^{-j_k t} (1-\cos(j_k p)) \\
            \geq & -\sum_{k=1}^{\infty} (e^{-j_k t }- e^{-j_k\gamma}) +  \sum_{k=1}^{\infty} e^{-j_k (\gamma + 1)} (1-\cos(j_k p)). \\
\end{split}
\end{equation}
If we define 
\begin{equation}\delta(t) = \sum_{k=1}^{\infty} (e^{-j_k t} - e^{-j_k \gamma}),\end{equation}
and
\begin{equation}\label{d_p} d(p) = \sum_{k=1}^{\infty} e^{-j_k (\gamma + 1)} (1-\cos(j_k p)),  \end{equation}
($\delta(t)$ is not to be confused with the Dirac mass) the above bound can be rewritten in
a simpler form:
\begin{equation}\label{esti_global}
|f(a)-1|\geq d(Re(a)) - \delta(-Im(a))~{\rm with}~a\in \h, Im(a)>-(\gamma+1).
\end{equation} 

Noticing that $t\mapsto\delta(t)$ is a strictly decreasing smooth function of $t$ satisfying $\delta(\gamma)=0$ and that
$d(p)$ is clearly always non-negative, on can see that $-Im(a)\in(\gamma,\gamma+1)$ which
entails $|f(a)-1|\geq 0 - \delta(-Im(a)) > 0$,
that is, $f-1$ has no zeros in $\{z: -Im(z)\in(\gamma,\gamma+1)\}$.

{\it Remark}: Obviously, the choice of $1$ in $\gamma+1$ is arbitrary.
We could have chosen $\gamma+T$ for any $T>0$, without any change
to what has followed. By doing this and letting $T\rightarrow+\infty$ one can recover the well-known fact
that \textit{all the roots of $f-1$ have imaginary parts of at least $-\gamma$.} Moreover, since 
$j_1 /j_2$ is not rational, $1-\cos(j_1p)$ and $1-\cos(j_2p)$ are both zero only if
$p=0$, and hence $d(p)>0$ for $p\neq 0$.
It follows that \textit{all roots of $f-1$ except $a_0$ have imaginary parts strictly less than $-\gamma$},
which is also well-known. In the following, we shall fix the choice of $d(p)$ as in (\ref{d_p}).

To analyze better $f$ and $F$, one should begin with $d(p)$, which actually measures how 
the function $f$ around $z=p-i\gamma$ approximates $f$ around $-i\gamma$. 
To state this more precisely, let us fix a radius 
\begin{equation}\label{r}r\in(0,\min\{ \pi/2j_1, \gamma \}),\end{equation} 
such that in the closed ball (rectangle) $\overline{R}(-i\gamma,r)$, $f(z)-1$
has only $-i\gamma$ as a zero (since $-i\gamma$ is a simple zero of $f$, this is always possible), and note 
\begin{equation}\label{m}m=\inf\{|f(z)-1|: z\in \partial R(a_0, r) \cup \partial R(a_0, r/2)\},\end{equation}
which is positive. 
Now let us estimate $f(p+z)-f(z)$ for $z\in \partial R(a_0, r) \cup \partial R(a_0, r/2)$.
After expanding and some algebra, one easily reaches: 
\begin{equation}
\begin{split}
&|f(p+z)-f(z)| \\ 
\leq  & \sum_{k=1}^\infty e^{j_k Im(z)}|e^{-ij_k(p+Re(z))}-e^{-ij_kRe(z)}|\\
\leq  & \sum_k e^{-j_k (\gamma-r)}|e^{-j_kp}-1|.
\end{split}
\end{equation}

Clearly we can find $K$ such that $\sum_{k>K}$ of the above sum is smaller
than $m/4$. For the finite terms left, it should be noticed that $d(p)<d_0$ implies 
$1-\cos(j_k p)<d_0e^{j_k\gamma}\leq d_0e^{j_K\gamma}$. So, as $d_0\rightarrow 0$,
$j_kp \text{ mod } 2\pi \rightarrow 0$ for all $k\leq K$, which then implies $\sum_{k\leq K}\rightarrow 0.$
 In particular, we choose and fix $d_0>0$ so that 
\begin{enumerate}
\item $d(p)\leq d_0$ entails $\sum_{k\leq K}<m/4$ and
hence \begin{equation}\label{eq_approx}
|f(p+z)-f(z)|<m/2, z\in \partial R(a_0, r) \cup \partial R(a_0, r/2).
\end{equation}
\item $d(p)\leq d_0$ entails  $|j_1 p\text{ mod } 2\pi| < j_1 r/4$, {\it i.e.},
$$\exists m\in\Z,|p\text{ mod } 2\pi m/j_1| < r/4.$$
\end{enumerate}

The estimate (\ref{eq_approx}) together with the definition of $m$ (\ref{m}) implies
 $|f(p+w)- f(w)|<|f(w)-1|$ for $w\in\partial R(a_0,r)$ so, by Rouch\'e's theorem,
$f(p+z)-1$ and $f(z)-1$ have the same number of
zeros (with multiplicity) in $R(a_0,r)$, that is, equivalently, \textit{$f-1$ has exactly one
zero in the ball $R(p+a_0,r)$}. The same argument applied to $r/2$ tells furthermore
 that this zero is actually in $R(p+a_0,r/2)$.
 
We now consider 
\begin{equation}\label{U}
U=\bigcup_{p: d(p)\leq d_0} R(a_0+p, r),\end{equation}
which, by property (2) of $d_0$, can be rewritten as $U=\cup_{l\in\Z} U_l$ where 
$$U_l = \bigcup \{R(a_0+p, r): p\in\R , d(p)\leq d_0,  |p - 2\pi l / j_1|\leq r/4\}.$$
Let us look more closely at an individual $U_l$ which is not empty (it can be so \textit{a priori}). The set 
$I_l = \{p: d(p)\leq d_0,  |p - 2\pi l / j_1| \leq r/4\}$ is clearly closed and bounded, so let $s_l,t_l$ be its minimum
 and maximum respectively. Hence $s_l \geq 2\pi l / j_1 - r/4$, $t_l \leq 2\pi l/ j_1 + r/4$.
Then it is easy to see that 
\begin{equation}\label{u_l}
U_l \subset \{ z : Re(z)\in (s_l-r, t_l+r), |Im(z)+\gamma|<r\} \subset R(s_l+a_0, r)\cup R(t_l+a_0, r) \subset U_l.
\end{equation}
The second inclusion comes from $t_l-s_l\leq r/2<2r$. All terms are therefore equal.
Moreover, if $n>l$ and $U_n\neq\emptyset$, then $(s_n-r) - (t_l + r) \geq 2\pi / j_1 - 5r/2 >
( 2\pi - 5\pi/4) / j_1 =: P>0$. Hence $ (s_l-r, t_l+r)\cap(s_n-r, t_n+r)=\emptyset$ (they are seperated by at least $P$ indeed), so it can be concluded
that  \textit{$\{U_l\}$ is a disjoint family of rectangles whose union is $U$}.\\

We now take $U_l$ nonempty and look at the poles of $F$ in it. We know from property (1) of $d_0$ (applied to $s_l$) that $F$ has exactly one pole $z$ in $R(s_l, r)$, and, furthermore,
that $z\in R(s_l+a_0, r/2) \subset R(t_l+a_0, r)$ since $t_l-s_l\leq r/2$. So 
$z$ is also the only (simple) pole in $R(t_l+a_0, r)$ (again it is in $R(t_l+a_0, r/2)$ actually). 
As $U_l=  R(s_l+a_0, r)\cup R(t_l+a_0, r)$, we can conclude
that \textit{$U_l$ contains exactly one (simple) pole of $F$, which we note $a_l$} (this way, $a_0$ coincides with our previous
 definition). This $a_l$ can be found and defined if and only if 
$U_l$ is nonempty; so by abuse of language, whenever we say $\forall l\in\Z$ or 
write $\sum_l k(a_l)$ ($k(a_l)$ being a term involving $a_l$), it is meant to be restricted to all such $l$'s.

Since $a_l\in U_l$, and neighboring $U_l$'s are separated by at least $P$,
\begin{equation}\label{equispacing}
Re(a_l) - Re(a_n) \geq (l-n)P, \text{  } (l > n),
\end{equation}
 and, hence, $|Re(a_l)| \geq |l|P$ for any $l\in\Z$. This gives control on its location.
  We can also estimate its residue by Cauchy's formula:
 $$\mathrm{Res}(F;a_l) = \frac{1}{2\pi i} \int_{\partial U_l} \frac{f(z)}{1-f(z)} dz. $$
 Indeed, whenever $z\in \partial U_l \subset \partial R(s_l+a_0, r)\cup \partial R(t_l+a_0, r),$
 we know from property (1) of $d_0$ that $d(s_l)\leq d_0, d(t_m)\leq d_0\Rightarrow |1 / (f(z)-1)|< 2/m $,
 hence $|f(z)/(1-f(z))|< 2f(-i(\gamma-r))/m$. The length of the integration path (the boundary of a rectangle)
  being always no larger than $4r + 4\pi / j_1$, we have $|\mathrm{Res}(F;a_l)|\leq M(=2f(-i(\gamma-r))(4r + 4\pi / j_1)/m)$
 whatever $l$. Since $a_l$ is the only simple pole in $U_l$, we can now say that
\begin{equation}\label{principal}
\forall l, z\in U_l \Rightarrow F(z) = \frac{c_l}{z-a_l}+ \text{ holomorphic term, } |c_l|\leq M.
\end{equation}

It is now possible to define the series
\begin{equation}\label{series}
E(z) =\sum_{l\in\Z} \frac{c_l}{a_l (z-a_l) }.
\end{equation}
The sum is over all $l$ such that $U_l\neq 0$. 
The convergence issue is now a standard exercise: since $|c_l|\leq M$ and $|Re(a_l)|\geq  P|l|$,
 \textit{the sum converges absolutely and uniformly in any compact subset of $\C$.}
It follows that $E(z)$ is a meromorphic function defined on $\C$. So we can define the difference
\begin{equation}\label{g_eta}
G(z) =F(z)/z - E(z), z\in\h.
\end{equation}
By construction, $E(z)$ removes the poles of $F(z)/z$ in $U$, so
$G$ is holomorphic in $U$. However, the important question is to understand if $E$ has removed the poles with the ``most negative" imaginary parts of $F(z)/z$.

The answer is positive. To see this, let us fix a positive $\eta$ such that
\begin{equation}
\eta < r, \delta(\gamma-\eta)<d_0/2.
\end{equation}
Immediately, (\ref{esti_global}) implies
that for $z=p-it$ with $p,t\in\R $,
\begin{equation}\label{f_good_outside}
d(p)\geq d_0, t\in (\gamma-\eta,\gamma+1) \Rightarrow |f(z)-1| \geq d_0 - d_0/2 = d_0/2.
\end{equation}
In particular, $f(z)-1=0$ and $t=-Im(z) > \gamma-\eta$ implies $d(p) < d_0$. 
Since $Re(z)=Re(p+a_0)$ and $|Im(z-(p+a_0))|<\eta<r$, we have 
$z\in R(p+a_0,r)\subset U$(recall the definition of $U$, (\ref{U})) and hence $z=a_l$ for some $l$. By construction,
$E(z)$ removes the singularity of $F(z)/z$ at $z=a_l$.
We have therefore shown that \textit{$G(z)$ has no pole whose
imaginary part is more negative than} $-\gamma+\eta$.\\

A closer attention should now be paid to this
observation. In fact, we can say that
\textit{$G \sim O(|z|)$ in the region $\Omega = \{z: Im(z)\in (-\gamma-1,-\gamma+\eta)\}$.}
To show this claim, 
let us first give a useful estimate of $E(z)$.
Let $z\notin U$. Then, for any $a_l$, we know that $a_l \in R(p+a_0,r/2)$ for $p=s_l$ or $t_l$.
In either case, we have $\norm{z-(p+a_0)}\geq r$ hence $|z-a_l| \geq \norm{z-a_l} \geq r/2$. So,
$$\mathrm{dist}(z,\{a_l:l\in\Z\})>r/2=\delta.$$
We can then estimate $E(z)$ from this fact alone, {\it i.e.} we can show that
\begin{equation}\label{esti_E_0}
\forall\delta>0, \forall z, \mathrm{dist}(z,\{a_l:l\in\Z\})>\delta\Rightarrow |E(z)|\leq
\sum_l  |c_l||a_l (z-a_l)|^{-1} \sim O(|z|).
\end{equation}
Indeed, \begin{equation}\label{sum_term}
\vert\frac{1}{a_l (z-a_l)} \vert\leq \frac{|z|+|a_l|}{\vert a_l(z-a_l)^2 \vert}  \leq 
|z-a_l|^{-2}(|z|/(\gamma-r)+1) = O(|z|) |z-a_l|^{-2},
\end{equation}
since $|a_l|\geq \gamma-r$. Thus
 $|E(z)|\leq (|z|/(\gamma-\eta)+1) \sum_l |z-a_l|^{-2}$.
Now, assume that $p=Re(z)\in[Re(a_{l-1}),Re(a_l))$. Thus
it follows from (\ref{equispacing}) that $|z-a_{l+k}|\geq Pk$
and that $|z-a_{l-1-k}|\geq Pk$, $k=1,2,\dots$. Hence
$\sum_l |z-a_l|^{-2} \leq 2 (\delta^{-2} + P^{-2}\sum_{k=1}^{\infty} k^{-2}) \sim O(1)$
and the bound depends only on $\delta$.
So we conclude, in particular, that 
$$
|E(z)| \leq  \sum_l  |c_l||a_l (z-a_l)|^{-1} \sim O(|z|), z\notin U.
$$
 We shall now show the above claim,
\textit{$G \sim O(|z|)$ in the region $\Omega = \{z: Im(z)\in (-\gamma-1,-\gamma+\eta)\}$},
 by splitting the problem into two cases: (1) $z\in \Omega\setminus U$, and (2) $z\in U\cap \Omega$.

\textit{Case (1)}: Assume $z\in \Omega\setminus U$. We know that $d(Re(z))\geq d_0$ (otherwise $z\in R(Re(z)+a_0,r)\subset U$);
(\ref{f_good_outside}) implies that $|f(z)-1|>d_0/2$. Since $|f(z)|$ is bounded in the region $\Omega$, and $0$ is far from it, 
$|F(z)/z|=|f(z)/(z(f(z)-1))|\sim O(1)$ is obvious.
Adding the estimate of $E(z)$ established above, this closes case (1).

\textit{Case (2)}: Let us recall (\ref{u_l}):
$$U_l = \{ z : Re(z)\in (s_l-r, t_l+r), |Im(z)+\gamma|<r\} = R(s_l+a_0, r)\cup R(t_l+a_0, r).$$
So, $w\in \partial U_l \Rightarrow |w|\sim \Theta(|l|+1)$. 
Hence (\ref{esti_E_0}) implies, in particular, that there is
a constant $C$ such that $w\in \partial U_l \Rightarrow |E(w)| \sim O(l)$.
Recall also that $|F|\sim O(1)$ on $\partial U_l$ ({\it c.f.} the  estimation
of (\ref{principal})), and so does $|F(z)/z|$. So $|G|\sim  O(|l|+1)$ on $\partial U_l$.
Therefore, by maximal modulus principle ($G$ being holomorphic in $U_l$), 
$|G| \sim  O(|l|+1) \sim O(|z|)$ for $z\in \cup_l U_l = U$. This is in fact
a stronger claim than required by case (2).\\

In summary, we have shown that $|G| \sim O(|z|)$ on 
$\Omega = \{z: Im(z)\in (-\gamma-1,-\gamma+\eta)\}$. It is easier
to show that $|G| \sim O(|z|)$ for $\{z: Im(z)\leq -\gamma-1\}$. Basically one has to use the fact that
since we are at least $1$ away from all the poles, $E(z)\sim O(|z|)$ (by (\ref{esti_E_0})).
So $|G| \sim O(|z|)$ on  $\{z: Im(z) < -\gamma+\eta \}$.
This property makes $G$ a exponentially negligible contribution to the asymptotic behavior of $\int S$, which
is the inverse Fourier transform of $F(z)/iz$. So, the major contribution is isolated and corresponds 
to $E$. We shall discuss these points in details in the next subsection.

\subsection{Back to real}

We now use the decomposition
$$F(z)/z = E(z)+ G(z), F(z) = zE(z)+ zG(z)=: e(z)+g(z),$$
to study the asymptotic behavior of $S$. To keep the distribution
theory machinery as light as possible, 
we choose a real smooth test function $\zeta(x)$ compactly supported in $\R^+$. Let $\phi(z)$ be its 
holomorphic Fourier transform
$$ \phi(z)=\int_\R \zeta(x)e^{-izx}dx,$$ which is an entire function. It has the following
decay property: 
\begin{equation}\label{decay_phi}
\forall m\in\N, \forall T\in\R, \sup_{t\leq T, p\in\R} |\phi(p-it) p^m| <\infty .
\end{equation}
This can be routinely checked by bounding $(d^m/dx^m) (e^{tx}\zeta(x))$ for all $t\leq T$
and then performing Fourier transforms.
Let us call $\zeta_A(x)= \zeta(x-A), A\geq 0$. Hence $\phi(z)e^{-iAz}$ is the 
Fourier transform of $\zeta_A$. We need to evaluate the quantity 
$$ \langle S, \zeta_A \rangle .$$

Let us fix $t>\gamma$, then elementary calculations 
yield $ \mathcal{F}[e^{tx} \zeta_A(x)] = \phi(p+it) e^{-iA(p+it)}$. 
Recall that $F(p-it)$ is the Fourier transform of $S_t = e^{-tx} S$.
One therefore obtains, by Fourier transform:
\begin{equation}
\begin{split}
\langle S, \zeta_A \rangle =& \langle e^{-tx} S, e^{tx} \zeta_A \rangle = e^{At}\int_\R 
F(p-it) \phi(p + it) e^{-iAp} dp \\
= & \int_\R dp \overline{\phi(p + it)e^{-iA(p+it)}}(g(p-it) +e(p-it)) \\
= & \int_{Im(z)=-t} dz \overline{\phi(\overline{z}) e^{-iA\overline{z}}}g(z)+ \int_{Im(z)=-t} dz\overline{\phi(\overline{z}) e^{-iA\overline{z}}}e(z) \\
= & I(A) + II(A). \\
\end{split}
\end{equation} 
All integrals make sense because of (\ref{decay_phi}) and thanks to the fact that $g(p-it)$ and $e(p-it)$ 
are both $O(|p^2|+1)$.\\

The term $I(A)$ is related with $g(z)$, which is holomorphic and has a $O(|z^2|)$ growth 
inside $\{Im(z)<-\gamma+\eta\}$. So fixing $t_1 \in (\gamma-\eta, \gamma)$, we have:
\begin{equation}\label{path_change}
 I(A) = \int_{Im(z)=-t_1} \overline{\phi(\overline{z}) e^{-iA\overline{z}}}g(z) dz .
\end{equation}
The function $\varphi(z):=  \overline{\phi(\overline{z})e^{-iA\overline{z}}}g(z)$ is indeed 
holomorphic in $\{Im(z)<-\gamma+\eta\}$. So
for $R>0$, Cauchy's theorem leads to:
$$
(\int_{-R-it_1}^{+R-it_1} dz - \int_{-R-it}^{+R-it}dz)\varphi(z) = (\int_{-R-it_1}^{-R-it}dz - 
\int_{R-it_1}^{R-it}dz)\varphi(z).
$$
As $R\rightarrow+\infty$, the right hand side tends to zero due to (\ref{decay_phi}), and the left hand side
tend to $- I(A) + \int_{Im(z)=-t_1} dz \varphi(z)$. So it vanishes and this establishes
(\ref{path_change}). Rewritten in the real variable $p$, this gives :
$$ I(A) = e^{At_1} \int_\R  \overline{\phi(p+it_1)} e^{iAp} g(p-it_1)dp.$$
The absolute value of the integral (without the $e^{At_1}$ prefactor) can be estimated by $\int_\R  |\phi(p+it_1)g(p-it_1)|dp$, which
is a finite constant, so we have:
\begin{equation}\label{i(A)}
|I(A)| = O(e^{At_1}), \gamma >t_1 > \gamma-\eta.
\end{equation}\\
 
Let us now turn to $II(A)=\int_{Im(z)=-t} dz\psi(z) zE(z),$ where 
$\psi_A(z):= \overline{\phi(\overline{z})e^{-iA\overline{z}}}$
and $E(z)$ is defined by the series: $$ E(z)=\sum_l \frac{c_l}{a_l (z-a_l)}.$$
This infinite sum can be interchanged with the integral in  $II(A)$.
Indeed, since $Im(z)=t\Rightarrow \mathrm{dist}(z,\{a_l:l\in\Z\})\geq t - \gamma >0$, by (\ref{esti_E_0}), 
$\sum_l \vert c_l /(a_l (z-a_l))\vert \sim O|z|$, so (\ref{decay_phi}) implies
that $$\int_{Im(z)=-t} dz |\psi(z) z|\sum_l \vert\frac{c_l}{a_l (z-a_l)}\vert \sim\int_\R |\phi(p+it)|O(|p-it|^2)e^{At}< +\infty.$$ So, by Fubini's theorem
(or dominated convergence theorem), the sum and the integral can be exchanged:
\begin{equation}\label{echange}
II(A)=\sum_l \int_{Im(z)=-t}  \frac{z c_l}{a_l (z-a_l)}\psi_A(z) dz=: \sum_l Q_l(A),\end{equation}
where $$Q_l(A):=  \int_{Im(z)=-t}  \frac{z c_l}{a_l (z-a_l)}\psi_A(z) dz = 
\int_{Im(z)=-t}  \frac{z c_l}{a_l (z-a_l)} \overline{\phi(\overline{z}) e^{-iA\overline{z} } } dz$$
can be explicitly calculated by Cauchy's residue formula for $A>0$ (the region in which we are interested).
The residue of the pole $a_l$ is indeed the only
contribution:
\begin{equation}\label{q_l}
 Q_l(A) = 2\pi i c_l \overline{\phi(\overline{a_l}) e^{-iA\overline{a_l}}} , A>0.
\end{equation}
This is a standard exercise in residue calculus (see the appendix).

Now we proceed to analyze $Q_l(A)$ in more details. To this end, let us write $a_l=p_l-it_l, (p_l,t_l\in\R)$, and notice that:
\begin{enumerate}
\item[-] $c_l$ is a nonzero coefficient, independent of $A$, and bounded for all $l$'s (\ref{principal}).
\item[-] $\overline{\phi(\overline{a_l})}$ is another coefficient independent of $A$.
 By (\ref{decay_phi}), for any $m\in\N$, $|\phi(\overline{a_l})|=|\phi(p_l+i t_l)|\sim O(p_l^{-m})\sim O(l^{-m})$ by (\ref{equispacing}). Combining the two points made so far, we have:
\begin{equation}\label{sum_k}
\sum_l |k_l| := \sum_l  2\pi |c_l \phi(\overline{a_l})| < +\infty.
\end{equation} 
\item[-] $\overline{e^{-iA\overline{a_l}}} = \exp({t_lA})\exp({i p_lA})$. This means that $Q_l(A)$ is exponentially growing
at speed $t_l$ and oscillates at the (angular) frequency $p_l$.
\end{enumerate}
In particular, we know that $p_0=0$ and $t_0=\gamma > t_l$ for any $l\neq0$. It is therefore
obvious that $$Q_0(A) = k_0 e^{\gamma A}$$ is the
dominating term in $II(A)$ when $A$ is large, unless one chooses a bad $\zeta$ such that $\phi(i\gamma)=0$.
This can easily be avoided if, for example, we require $\zeta$ to be nonnegative and nonzero. Then, 
$\phi(i\gamma)=\int_R \zeta(x)e^{\gamma x}dx$ is real and positive, so the coefficient $2\pi i c_0 \overline{\phi(i\gamma)}$ is actually also real and positive
since we can calculate, from the definition, that $c_0= -i(\sum_k j_k e^{-j_k \gamma})$. In the following we assume to be in this case.
Precisely, we need to show that
\begin{equation}\label{main_II}
\lim_{A\rightarrow+\infty}\frac{II(A)}{Q_0(A)} = 1. 
\end{equation}
\textit{Textbook proof}
For any $\epsilon>0$, by (\ref{sum_k}), one cas find $L>0$ such that:
$$\sum_{|l|>L} |k_l|< k_0\epsilon.$$
This entails that, for all $A$, 
$$ \sum_{|l|>L} |Q_l(A)|/|Q_0(A)| = \sum_{|l|>L} e^{(t_l- \gamma) A}k_l/k_0 < 
\sum_{|l|>L} |k_l|/|k_0| < \epsilon .$$
Next, As $t_l<\gamma$ for any $|l|\leq L, l\neq 0$, 
$$\sum_{|l|\leq L, l\neq 0} |Q_l(A)|/|Q_0(A)| = \sum_{|l|\leq L, l\neq 0} k_l/k_0 e^{(t_l-\gamma) A} 
\rightarrow 0 $$ for $A\rightarrow+\infty$. Hence
$$\limsup_{A\rightarrow+\infty} \sum_{l\neq 0} |Q_l(A)|/|Q_0(A)|\leq \epsilon.$$
This is true for all $\epsilon>0$, so $\sum_{l\neq 0} |Q_l(A)|/|Q_0(A)|\rightarrow 0$ as
$A\rightarrow+\infty$, which is equivalent to (\ref{main_II}). $\square$

Since $I(A)\sim O(e^{t_1A}) \sim o(e^{\gamma A}) \sim o(Q_0(A))$ ({\it c.f.} (\ref{i(A)}) and note that $t_1<\gamma$), the above equation
implies (recall that $I(A)+II(A)=\langle S, \zeta_A \rangle$):
\begin{equation}\label{main}
\lim_{A\rightarrow+\infty}\frac{\langle S, \zeta_A \rangle}{Q_0(A)} = 1, (Q_0(A) = k_0 e^{\gamma A})
\end{equation}
that is, \textbf{the goal of this study}.

\section{Discussions}\label{discussion}
The crucial quantity
$$\langle S, \zeta_A \rangle$$ 
reflects the asymptotic behavior of $S$ for big $A$, in the specific way. Let us take
$\zeta$ to be an approximation of the Dirac mass $\delta(x-a)$.
In this case, $\langle S, \zeta_A \rangle$ measures 
the number of states ``having area $A+a$'' with an error due to the smoothing out, just
 as in any real life physical process/detector. 
Moreover, the ``resolution'' of the measure is ``of $O(1)$'' when $A$ gets big, that is, the measure $\langle S, \zeta_A \rangle$ 
 is only dependent of $S$ restricted to $(A+a-\delta,A+a+\delta)$, where $\delta$ depends on $\zeta_A$ (the detector)
  but not of $A$.
  The result (\ref{main}) thus means that the outcome of this measure is $k_0 e^{\gamma A}(1+o(1))$, where $k_0$
  depends on the detector $\zeta$ (its offset, its resolution, etc). Taking logarithm: 
  \begin{equation}
  \log \langle S, \zeta_A \rangle = \gamma A + o(1)+ \log k_0.
  \end{equation} 
  This is a \textit{\textit{Bekestein-Hawking like entropy behavior}}. (Provided that 
  the Immirzi parameter is tuned to produce the same slope.)\\

   The energy of a black hole is related with its area $A$ through $E\sim \sqrt{A}$, and its
  temperature is given by $T\sim E^{-1}$. It is therefore easy to see that the area change caused by the emission of  one quantum is of the Planck area order, regardless of the size of the black hole. Thus
  even for macroscopic black holes, the relevant area interval (on which we count microscopic states to calculate the entropy) is $[A-a, A]$ with $a\sim O(1)$ as $A\rightarrow\infty$. To illustrate why this observation is essential, let us consider the following (hypothetical) scenario:  \textit{What would happen if $j_k$ were all integers, say $j_k=k,k\in\N$? } In this cas, \textit{ (\ref{main}) would no longer be valid.} Indeed,
  $S$ would peak only at integral values, thus $\langle S, \zeta_A\rangle $ would be highly periodic. The Hawking radiation spectrum would be totally different from the usual semi-classical theory, even for macroscopic black holes. 
  In our Fourier-analytic picture, the proof of (\ref{main_II}) would break down: $f(z)$ and $F(z)$ would be both 
  periodic (with period $2\pi$), therefore there would be infinitely many $Q_l(A)$ having the same growth rate than $Q_0(A)$.

This is in sharp contrast with ordinary thermodynamics. In this case, when counting microscopic states having energy in the interval $[E-\delta E,E]$, $\delta E$ increases as $E$ does (then one divides the result by $\delta E$ to normalize). For example, let us consider a system of $N$ independent spins of moment $\mu$ in a magnetic field $B$. The allowed energy levels are of the form $k\mu B$ with $k$ belonging to a subset of integers. Yet it is possible to use continuum analysis, {\it e.g.} the Stirling formula, to study its macroscopic behavior. Correspondingly, in the Fourier picture,
  one is allowed to neglect contributions other than $Q_0$, exactly because the test function $\zeta$, in order to fit in $[E-\delta E, E]$, is translated and \textit{dilated}. So $\phi$(the Fourier transform of $\zeta$) gets more and more localized, making the poles with periodic contributions less and 
  less relevant. Approaches like this
  are not \textit{a priori} suitable for the study of black hole thermodynamics. Since they are common practice in ordinary thermodynamics, special
   attention must be paid when one wishes to apply usual tools to understand black hole thermodynamics. This is precisely why
    we have performed the analysis given in this note. \\

The analysis of this note is qualitative and, we believe, far from optimal. For example,
 it would be interesting to estimate the error term in (\ref{main}). To this end, it would be necessary to
 understand better the behavior of the poles of $F$. Another possible technical improvement would be to weaken the smoothness  constraints on the test function.  \\

This study gives anyway strong arguments against any revival of the periodic structure of the entropy. 
Any possible way to revive this ``low mass" phenomenon would be extremely unnatural. Although this might look disappointing from the phenomenological viewpoint (see, {\it e.g.}, \cite{bar}), this ensures a correct ``low temperature" behavior, in agreement with the Bekenstein-Haking derivation which is somehow unavoidable at the classical level. In our opinion, this result makes even stronger the case for the LQG computation of black hole entropy.\\

A final remark should be made about the various prescriptions (summing schemes) found in the literature . In \cite{diaz1}, the distribution $S$ studied in this note takes into account the degeneracy caused by the Pell equation and by the re-ordering (r-degeneracy). On this point, almost all prescriptions agree one with the other.
However, they differ in the way they treat the  \textit{projection constraint}. In fact, for the prescriptions that mainly ignore this constraint, the analysis of
 our study can be applied mostly \textit{verbatim}, leading to the same conclusion (though the growth rate $\gamma$ will
 vary from one prescription to another). On the other hand, a rigorous treatment of the projection constraint introduces intricate singularities for $F(z)$, as it can be seen in \cite{recent_bh}. We have not attempted, in this study,
 to give an exhaustive treatment of the problem. We leave this question opened for a future article. Nevertheless, we believe that the result should most likely remain
 unchanged.

\section*{Appendix}
\textit{Proof of (\ref{q_l})}
 Let us choose $R>0$ and the consider the half-disk
contour integral $$(\int_{-R-it}^ {R-it} + \int_{\partial B^+(-it,R)}) z c_l/(a_l (z-a_l))\psi_A(z) dz,$$
where $\partial B^+(-it,R) =\{-it + Re^{i\theta}: \theta\in[0,\pi] \} $. 
By using the usual residue formula, one can show that, when $R$ is big enough so that $a_l\in B(-it,R)$, this integral is equal to
 $$2\pi i \mathrm{Res}( \frac{z c_l}{a_l (z-a_l)}\psi_A(z); a_l).$$
It suffices to show that
 $\int_{\partial B^{+}(-it,R)} \rightarrow 0$ when $R\rightarrow +\infty$. For this, we
 parametrize with the angle $\theta$:
 $$\int_{\partial B^{+}(-it,R)} \frac{z c_l}{a_l (z-a_l)}\psi_A(z) dz = i \int_0^{\pi} \frac{z(\theta) c_l}{a_l (z(\theta)-a_l)}\psi_A(z(\theta))   Rd\theta, $$
 where $z(\theta)= z(\theta;R)=-it+Re^{i\theta}$, and estimate the integrand's factors. When $R\rightarrow +\infty$,
  $z(\theta) c_l/ (a_l (z(\theta)-a_l)) \rightarrow c_l$ uniformly for $\theta\in[0,\pi]$, in particular
  this factor $\sim O(1)$ (uniformly in $\theta$) when $R$ is large.
  
  Now, let us consider $|R\psi_A(z)|= R|\phi(\overline{z})||e^{-iA\overline{z}}|
  = R|\phi(\overline{z})| e^{A(t-R\sin\theta)}$. It is uniformly bounded for $R>0,\theta\in[0,\pi]$.
  Indeed, for any $R>0, \theta\in[0,\pi]$, $Im(\overline{z(\theta)})\leq t$ so, by (\ref{decay_phi}),
  $|\phi(\overline{z(\theta)})|\sim O((R|\cos\theta|+1)^{-1})$ for all $(R,\theta)$. On the other hand,
  $|e^{-iA\overline{z(\theta)}}| = e^{A(t-R\sin\theta)} \sim O(e^{-AR\sin\theta})$
  for all $R,\theta$. So $|R\psi_A(z)| \sim O (R (R|\cos\theta|+1)^{-1}e^{-AR\sin\theta})$ for all $(R,\theta)$.
  Now we split into two cases: $\theta\in[\pi/3, 2\pi/3]\Rightarrow |R\psi_A(z)| \sim R (R|\cos\theta|+1)^{-1} \sim O(R\times R^{-1}) \sim O(1)$;
   $\theta\in[0,\pi/3]\cup [2\pi/3,\pi]\Rightarrow e^{-AR\sin\theta}\leq (A\sin\theta)^{-1} \sim O(1)$, hence
   $|R\psi_A(z)|\sim O(1)$ again. Hence $|R\psi_A(z)|$ is uniformly bounded.  
  Moreover, when $R\rightarrow\infty$, pointwisely, $e^{A(t-R\sin\theta)}\rightarrow 0$ 
  for any $\theta\in(0,\pi)$, {\it i.e.}, for almost every $\theta\in[0,\pi]$. Combining this with the estimate of
  $z(\theta) c_l/ (a_l (z(\theta)-a_l))$, we can apply the dominated convergence theorem to obtain
  $$\lim_{R\rightarrow+\infty} \int_0^{\pi} \frac{z(\theta) c_l}{a_l (z(\theta)-a_l)}\psi_A(z(\theta)) Rd\theta=0,$$
as desired. 
$\square$

\end{document}